# Nonlinear photoelectron emission from metal surfaces induced by short laser pulses. The effects of field enhancement by surface plasmons

Sándor Varró and Norbert Kroó

Research Institute for Solid State Physics and Optics of the Hungarian Academy of Sciences
H-1525 Budapest, P. O. Box 49, Hungary,
E-mail: varro.sandor@wigner.mta.hu, varro@mail.kfki.hu, phone: +36-1-392-2635

**Abstract.** Nonlinear electron emission processes induced by surface plasmon oscillations have been studied both experimentally and theoretically. The measured above threshold electron spectra extend up to high energies whose appearance cannot be explained solely by standard nonperturbative methods, which predict photon energy separated discrete energy line spectra with the known fast fall – plateau – cutoff envelope shape, even when taking the large field enhancement into account. The theoretical analysis of our data, based on the concept of plasmon-induced surface near-field effects gives a reasonably good explanation and qualitative agreement in the whole intensity range.

**Keywords:** surface plasmons, field enhancement, nonlinear surface photoelectric effect

**1. Introduction**

Surface plasmons (SPO-s) are wave-like motions of conduction electrons on a metal surface, coupled with photons, e.g. the ones exciting these waves. The energy of these SPO-s in this case is that of the exciting photons, but their wavelength is shorter. One of the main characteristics of the SPOs is their very large electromagnetic fields concentrated at the interfaces of metals and dielectrics (vacuum). To our knowledge, such an enhancement was first discussed by Fano [1] in 1938. Recently this phenomenon has become the subject of extensive research, because on the basis of it high-order nonlinear processes can be induced even at relatively moderate intensities of an incomig radiation which excites the SPOs [2-5]. In the case of a simple thick metal surface the general theory predicts the well known characteristic above threshold "fast fall – plateau – cutoff" envelope photoelectron shape, which consists of discrete spectral lines, repeated with steps of the exciting photon energy. It is, however, quite reasonable to expect, that due to the strongly nonlinear SPO induced near field effects, including the strong field enhancement, the photoelectron spectrum may have a different shape due to the mediating involvement of the resonantly excited SPO-s on the thin metal film. Consequently, in addition to the perturbative – non-perturbative theories, valid for metals, the elaboration of our special theoretical model [6] seems to be feasible, which incorporates the special features of the SPO excitation process.

In our experiments ~80fs Ti:Sa laser pulses (wavelength ~800nm, photon energy 1.5eV) were used to excite SPOs at a thin gold surface in the Kretschmann geometry, and the energy spectra of the generated photoelectrons have been analysed by the time-of-flight technique. In these emission processes the minimum four times of a plasmon energy is needed to overcome the work function (4.7eV) and deliberate the photo(plasmo) electrons. Relatively low laser intensities (in the range 2.70-23.1 . $10^8$ W/cm$^2$) have been used, and electrons of energies up to 40eV were observed. The model of [6] is capable to explain qualitatively these observations too, and the analytic formula, derived from this model, has proven to be in a quite good (qualitative) agreement with the experimental data. Furthermore, according to our recent theoretical analysis [7], in line with these observations based on the huge SPO field enhancement, it can not be excluded, that the emitted electrons come out from the surface in attosecond pulse trains, due to the interference of the higher order above-threshold de Broglie waves at the metal surface, even in the case of the used relatively moderate incoming laser intensities.

## 2. Experimental results

The aim of our experimental work has been to study and compare with theoretical predictions the spectral properties of SPO emitted electrons at these moderate laser intensities. The layout of the experimental set-up is shown in Fig 1a. The exciting light source was a frequency doubled Nd:YAG laser pumped long cavity Ti:Sa oscillator with a central wavelength of 800nm and a pulse length of about 80fs (after compression) with a repetition rate of 3.6MHz. The average power of the light beam was about 6-700mW. The laser pulses excited the SPO-s in air via a glass prism in the Kretschmann geometry. It was focused on the 50nm thick gold film, deposited on the glass prism, by a 18.5mm focal length lens (Fig.1a). The SPO exciting laser intensity was changed by moving this lens out of focus. The emitted photoelectrons were analyzed in vacuum ($<10^{-6}$Torr) by a time-of-flight spectrometer (Fig1b). The flight path of the electrons was 48.5cm, the entrance aperture 0.7cm and the acceptance angle 38°. In order to avoid frame overlapping of electrons coming from consecutive laser pulses at this high (3.6MHz) repetition rate an 5eV positive drift voltave was applied. The detector was a multichannel plate multiplier (Photonic Chevron MCP) and the its signals were recorded by a multichannel scaler with a time resolution of 100ps. The slope of the photocurrent vs. laser intensity curve, when plotted on double logarithmic scale was linear at the lower end of the intensity with a slope of 4, at higher intensities the slope went down to around 2 and below.

Six typical electron spectra are shown in Fig. 2 for relatively low intensities. The number of electron pulses per laser pulse were far below one, so no space charge effects had to be considered. First the laser intensity was varied in the ~59-200 MW/cm$^2$ range. For larger intensities the measured spectra become more structured, certainly because of possible heating effects or additional classical acceleration mechanisms. The count rate decreased significantly with increasing laser intensity. At the lower end of the laser intensity range the electron spectrum is peaked at around 5eV and the detailed analysis of this spectrum showes some periodic structure in it, indicating the presence of above-threshold 4th, 5th, 6th, etc. order currents induced by the SPO. At the upper end of the exciting laser intensity range the electron spectra are also peaked, but at maxima which are at higher energy and are shifted to higher electron energies with increasing laser intensity. The result of the Fourier transformation in these cases is a flat, structureless function. At intermediate laser intensities there is a continuous transition between the two cases presented in this figure. As seen in Fig.2 we have not found plateaus in the electron spectra in the experiments which were carried out with relatively long laser pulses, comparable with the life-time of the SPO-s at this photon energy.

## 3. Theoretical considerations

In the prism of index of refraction $n = 1.5$ the incoming p-polarized laser radiation of circular frequency $\omega_0$ represented by the electric field strength

$$\vec{E}_0 = (\vec{\varepsilon}_x \cos\theta_0 + \vec{\varepsilon}_z \sin\theta_0) | F_0 | \cos\left[\omega_0\left(t - n\frac{x\sin\theta_0 - z\cos\theta_0}{c}\right) + \varphi_0\right], \quad (1)$$

where $c$ is the velocity of light in vacuum, $\varphi_0$ is the phase of the complex amplitude $F_0$, and $\vec{\varepsilon}_x$ and $\vec{\varepsilon}_z$ are unit vectors being parallel and perpendicular to the metal surface, respectively. The numerical value $\varepsilon_2(\omega_0) = -\varepsilon_R + i\varepsilon_I = -25.82 + i1.63$ of the dielectric constant of the gold layer has been taken from Johnson and Christy [8]. With these parameter values, above the angle of total reflection $\theta_t = 41.81°$, at the critical angle of incidence $\theta_0 = \theta_c = 42.84° = \theta_t + 1.03°$ the amplitude of the reflected wave drops practically to zero in a narrow angular range of half width of about $0.4°$, and surface plasmon oscillations are generated at the metal-vacuum interface. In our experiments this 100% attenuated total reflection has been clearly observed. In our analytic calculations for the elliptically polarized electric field strength $\vec{E}_{sp}$ representing the SPOs [9] in vacuum (z >0) the following expression has been obtained

$$\vec{E}_{sp}(\vec{r},t) = g | F_0 | e^{-x/2L_{sp}} e^{-z/2l_z} [\vec{\varepsilon}_z \cos(\omega_0 t' + \varphi_0) - (\varepsilon_R - 1)^{-1/2} \vec{\varepsilon}_x \sin(\omega_0 t' + \varphi_0)], \quad (2a)$$

$$L_{sp} \equiv |2\,\mathrm{Im}[k_{sp}]|^{-1}, \quad l_z = (\varepsilon_R - 1)^{1/2} \lambda_0 / 4\pi, \quad t' \approx t - (n_1 \sin\theta_c)(x/c). \quad (2b)$$

In Eq. (7c) we have introduced the the 'propagation length' $L_{sp} \approx (\varepsilon_R - 1)^2 \lambda_0 / 2\pi\varepsilon_I$ of the SPO, which is $\sim 61.6 \times \lambda_0 = 49\mu m$ for $\varepsilon_R = 25.82$, $\varepsilon_I = 1.63$ and $\lambda_0 = c/\omega_0 = 795nm$. The dimensionless factor $g$ in the field amplitude in (2a) is about 12, thus the direct SPO intensity enhancement is on the order $g^2 \approx 150$ in the present case. If we take into account the geometrical field enhancement due to surface irregularities of spherical shape, then the maximum intensity enhancement factor was certainly about $\bar{g}^2 \approx 600$ in our experiments. It is seen in Eq. (8b), that in vacuum the SPO has an elliptically polarized electric field whose longitudinal $x$-component is smaller by a factor of $(\varepsilon_R - 1)^{1/2} \approx 5$ than the $z$-component perpendicular to the metal surface. Inside the metal, on the other hand, the longitudinal component is larger by this same factor than the perpendicular one, so, this latter is smaller by a factor of $(\varepsilon_R - 1) \approx 25$ to compare with its value in vacuum. The fields inside the metal exponentially drop to their $1/e$-value within the distance of $\lambda_0 / 31.3 \approx 25.4 nm$.

The Schrödinger equation of a metallic electron interacting with the SPOs reads

$$\hat{H}\psi(\vec{r},t) = i\hbar \frac{\partial \psi(\vec{r},t)}{\partial t}, \quad \hat{H} = \frac{1}{2m}\left(\hat{\vec{p}} + \frac{e}{c}\vec{A}_{sp}\right)^2 + V, \qquad (3)$$

where $\hat{\vec{p}} = -i\hbar\nabla$ is the electron's momentum operator in coordinate representation, $m$ is its mass, and $e$ is the elementary charge. The electric field strength $\vec{E}_{sp}$ given by Eq. (2a) is connected with the vector potential $\vec{A}_{sp}$ by the relation $\vec{E}_{sp} = -\partial \vec{A}_{sp}/\partial ct$. It is customary to use the Sommerfeld step potential model for the unperturbed electrons, thus in the Hamiltonian given by Eq. (3) we take $V(z) = -V_0$ for $z < 0$ and $V(z) = 0$ for $z > 0$, where $V_0 = E_F + W$ with $E_F = 5.51 eV$ and $W = 4.68 eV$ being the Fermi energy and the work function, respectively, for gold. Besides, we are allowed to arbitrarily set $\varphi_0 = 0$ for the initial phase, because possible carrier-envelope phase difference effects [10,18] are not relevant here, since in the experiments we have used relatively 'long' pulses containing more than fourty cycles. The electron transitions are induced at the metal-vacuum interface in a very narrow transverse spatial region, $|z| \ll \lambda_0$, thus it is justified to use the usual dipole approximation. Moreover, it is well-known [11] that the basic features of the surface photoelectric effect can be described by taking into account only the z-component of the electronic motion being perpendicular to the metal. This is because, at least according to the Sommerfeld model we are using, the electrons are moving freely along the metal surface, and in the photon (plasmon) absorption process only the z-component of the momentum changes significantly. In the dipole approximation and in one dimension Eq. (3) simplifies to the equation

$$[(\hat{p} + eA/c)^2 / 2m]\varphi = i\hbar\partial_t\varphi \quad , \qquad \hat{p} = -i\hbar\partial/\partial z, \qquad A = (c/\omega_0)\overline{g}F_0 \cos\omega_0 t,$$
(4)

where now $F_0$ denotes the amplitude of the z-component of the electric field strength representing the incoming laser field, and $\overline{g}^2 \approx 600$ is the total intensity enhancement factor, as has already been discussed above. The exact fundamental solutions of Eq. (4) are the nonrelativistic version of the Volkov states which have long been widely used in strong-field physics [12]. They are modulated plane waves of the form

$$\varphi_p(z,t) = N \exp[(i/\hbar)(pz - E_p t)] \exp[-i(\bar{g}\mu_0 pc/\hbar\omega_0)\sin\omega_0 t] e^{-if(t)}, \quad E_p = p^2/2m, \quad (5a)$$

$$\mu_0 \equiv eF_0/mc\omega_0 = 10^{-9} I^{1/2} \lambda_0, \quad f(t) \equiv (1/\hbar)\int dt (e^2/2mc^2) A(t). \quad (5b)$$

In Eq. (5b) we have introduced the dimensionless intensity parameter $\mu_0$ of the incoming laser radiation, whose numerical value can be easily calculated by the second equation in the defining equation where the intensity $I$ measured in $W/cm^2$ and the wavelength $\lambda_0$ is measured in $10^{-4} cm$. In the nonperturbative Keldish-type description [11-12] one would use the Volkov states, Eq. (5a), as final states and expands into a Fourier series the exponential with the sinusoidal modulation with the help of the Jacobi-Anger formula [13]

$$\exp[-i(\bar{g}\mu_0 p_n c/\hbar\omega_0)\sin\omega_0 t] = \sum_{k=-\infty}^{\infty} J_k(\bar{g}\mu_0 p_n c/\hbar\omega_0) e^{-\frac{i}{\hbar} n\hbar\omega_0 t}, \quad p_n = \sqrt{2m(-W + n\hbar\omega_0)}, \quad (6)$$

where $J_n(x)$ are ordinary Bessel functions of first kind of order $n$. As a result, the $n$-plasmon absorption probability would be proportional with $J_n^2(\bar{g}\mu_0 p_n c/\hbar\omega_0)$. At the lowest intensity $I = 2\times 10^8 W/cm^2$ we used in our experiments $\mu_0 \approx 10^{-5}$ and, even if we take into account the enhancement factor $\bar{g} = 2\times 12$, the argument of the Bessel function would be on the order of or less than 1, and, of course less than $n_0 = 4$, the minimum number of absorbed plasmons needed for the deliberation of the electrons from the metal. In this case, owing to the asymptotic behaviour of the Bessel functions $J_n^2(x) \approx [(x/2)^n/n!]^2$, the above-threshold peaks would exponentially drop, in sharp contrast to our experimental results. This means that on the basis of the standard nonperturbative description it is not possible to interpret the measured spectra, even if we take into account the considerable intensity enhancement of order 600, stemming from the SPO generation.

In order to have a satisfactory theoretical interpretation of our experimental results, as a next step in the analysis, we have applied the so-called 'laser-induced near-field' model which have long been introduced [14-16] by one of us (see also the more recent studies in [7], [9] and [6]). In this description the basic interaction leading to very high nonlinearities is caused by the collective velocity field of the oscillating electrons near the metal surface, within a layer of thickness smaller that the penetration depth $\delta$. Because the quasistatic velocity field is screened inside the metal, the thickness of the layer is taken as the Thomas-Fermi screening length $\delta_s = 1/k_{TF}$, where $k_{TF} = (6\pi n_e e^2/E_F)^{1/2}$. The concept of the laser-induced collective near-field can be illustrated by the following physical picture. The electric field component of the radiation field being perpendicular to the surface makes the electrons oscillate, but the ionic cores remain stationary. According to Newton's second equation, these electrons aquire an additional oscillatory displacement $\vec{\alpha}(t) = \vec{e}_z \bar{g} \mu_0 (\lambda_0 / 2\pi) \sin(\omega_0 t - \vec{k}_\parallel \cdot \vec{x}_i)$ along the z-direction, which is superimposed to their average position $\vec{x}_i$. The additional potential energy of a test electron (the electron to be freed, with position $\vec{r}$) stems from the attraction of the ionic cores, and from the repulsion of the (now oscillating) electrons; $V_i(\vec{r}) = e^2 / |\vec{x}_i + \vec{\alpha}(t) - \vec{r}| - e^2 / |\vec{x}_i - \vec{r}|$, where the index $i$ refers to the i-th surface electron (which is associated on average to the i-th ionic core). By summing up these contributions (in the continuum limit; $\sum_i V_i \to n_e \int d^3x$), we obtain an oscillating double-layer potential acting on the test electron. Inside the metal the effect of this quasi-static field of the oscillating electron layer is only essential down to the Thomas-Fermi screening length $\delta_s$, thus the z-integration is restricted over this thickness. An inner test electron approaching the boundary, periodically feels additional repulsion (negative excess charge) or attraction (positive excess charge), depending on the phase of the inducing electric field of the surface plasmon. Out of the metal (z > 0; in vacuum) a similar situation occurs, but in an opposite manner, because there is an amplitude jump $V_D - (-V_D) = 2V_D$ in this potential at crossing the metal-vacuum boundary (z = 0). The wave function of an electron will then obey the two Schrödinger equations

$$(\hat{p} + eA/c)^2 / 2m - V_0 - V_D \sin \omega_0 t) \Psi_I = i\hbar \partial_t \Psi_I \qquad (z < 0), \tag{7a}$$

$$(\hat{p} + eA/c)^2 / 2m + V_D \sin \omega_0 t) \Psi_{II} = i\hbar \partial_t \Psi_{II} \qquad (z > 0), \tag{7b}$$

where the subscript *I* refers to the interior region (metal) and *II* to the exterior region (vacuum), respectively. Following Refs. [14-16] the amplitude of the collective velocity field

$$V_D = 2\pi n_e e^2 \xi_0 \delta_s = \bar{g}\mu_0 (\omega_p / 4\omega_0)(\delta_s / \delta)(2mc^2), \quad V_D \approx \bar{g}\mu_0 \times 10^4 eV, \tag{8}$$

where the dimensionless parameter $\mu_0$ has already been defined in Eq. (5b), and $\delta$ denotes the skin-depth of the metal at $\omega_0$. Of course, the interaction with the SPO field in principle should still be taken into account, too, as in Eq. (4), but since its direct effect is much smaller than that of the induced collective velocity field, we have not displayed this direct interaction in Eqs. (7a-b). The time-averaged outgoing electron current components for which the momenta $p_n = [2m(n\hbar\omega_0 - W)]^{1/2}$ are real ($n \geq n_0 = 4$ in the present case), corresponding to $n$-order multiplasmon absorption, have been obtained from the Fourier expansion of $\Psi_I$ and $\Psi_{II}$. The unknown multiphoton reflection and transmission coefficients, $R_n$ and $T_n$, respectively, can be determined from the matching equations, i.e. from the continuity of the wave function, $\Psi_I(0,t) = \Psi_{II}(0,t)$ and of its spatial derivative, $\partial_z \Psi_I(0,t) = \partial_z \Psi_{II}(0,t)$, which relation must hold for arbitrary instants of time. The resulting two coupled infinite set of linear algebraic equations for $R_n$ and $T_n$ can be numerically solved without any particular difficulty, moreover in our earlier works [14-16] we have derived quite accurate analytic approximate formulas, too. According to these resuts, the current components normalized to the incoming current can be expressed as

$$j_t(n) = (p_n/q_0) \cdot |T_n|^2, \quad |T_n|^2 \approx J_n^2(a) \quad (n \geq n_0), \quad a \equiv 2V_D/\hbar\omega_0, \qquad (9)$$

where $q_0 = (2mE_F)^{1/2}$ is the average of the initial momenta (see also Ref. [6]). For instance, in case of $I = 2 \times 10^8 W/cm^2$ we have $2V_D \approx 11 eV$, and $a = 2V_D/\hbar\omega \approx 7$, where we have taken for the ratio $\delta_s/\delta = 2 \times 10^{-2}$, i.e. for $\delta = 22.5 nm$ the screening length is about $\delta_s \approx 0.4 nm$. This numerical example clearly shows that already at the lowest intensity used in the experiments the *new nonlinearity parameter a* introduced in Eq. (9) has a much larger value than the argument of the Bessel function in Eq. (6). On the basis of this remarkable quantitative difference, our theory based on introducing the laser-induced near-field is capable to account for the basic features of the measured electron spectra. As we have seen before, in the frame of the standard *nonperturbative* Volkov description there is no chance to interpret this experiment, and several earlier experiments (as has also been emphasized in Refs. [14-16], [7] and [17]). The theoretical results are summarized in Fig. 3. The calculated above-threshold electron spectra are shown here for four different values of the incoming laser intensity, for illustrational purposes. We note that these figures have been generated without any adjustments, only on the basis of the formula in Eq. (9). Besides the numerical values of the natural constants $c$, $\hbar$, $k$, $e$ and $m$, the experimental values of the optical constants of gold [8], its electron density $n_e = 5.9 \times 10^{22}/cm^3$, and the independently measured intensity of the laser have been used as input parameters for the calculation. Fig. 3a corresponds to the experimental curve labeled by "6" in Fig. 2. A comparison with the experimental curves has shown that, in spite of the relative simplicity of our theoretical model based on the effect of the laser-induced near-field at the metal-vacuum interface, the agreement is reasonable between experiment and theory. However, a qualitative discrepancy was found in higher laser intensity cases, where probably additional concurrent heating effects may have played some role. In the above model, these thermal and some other possible effects (e.g. due to resonant image potential states [19]) have not been taken into account.

In Figure 4 we illustrate a comparison of the intensity dependence of the total current in case of perturbation theory, and according to our new non-perturbative description. For the sidebands of larger indeces, the total current may be approximated by (if we neglect the $n$-dependence of the higher momenta)

$$j = \sum_n j_n \propto \sum_{n=5}^{\infty} J_n^2(a) = (1/2)\left[1 - J_0^2(a)\right] - \left[J_1^2(a) + J_2^2(a) + J_3^2(a) + J_4^2(a)\right]$$

, where the sum rule of the Bessel functions has been used; $\sum_{n=-\infty}^{\infty} J_n^2(a) = 1$. The $n=4$ term is relatively small in comparison with higher terms, due to the smallness of the prefactor in $\left[(n\hbar\omega_0 - W)/E_F\right]^{1/2} J_n^2(<a>)$ immediately close to threshold ($|4\hbar\omega_0 - W| << n\hbar\omega_0$). In the figure it is clearly seen that, though at even relatively low intensities there are quite many above-threshold electrons (as shown by see Figs. 3a-b), the sum of the currents follow the 4[th] power perturbative law. For larger intensities the 'degree of non-linearity' $n(I) = \partial \log j / \partial \log I$ strongly deviates from n=4, and the total current saturates, which is a naturally expected behaviour.

## 4. Conlusions

A platoless energy distribution of SPO mediated electron emission was found, when the pulse length of the SPO exciting laser was comparable with the life-time of the surface plasmons. The electron emission was found to step in already at relatively low laser intensities ($10^8$ W/cm$^2$) and the electron spectrum was found not to have a plato in contrast to expectations. In the Fourier transform of the energy spectra we have seen the discrete structure at low intensities, which comes out from our theory, too. On the other hand, the appearance of the large-energy electrons cannot be explained with standard non-perturbative theories, even if one takes into account the field enhancement due to the surface plasmons. If one takes some classical acceleration mechanism, one could perhaps explain the large energy part of the spectrum, but, needless to say, one cannot account for the appearance of the discrete structure with spacing proportional with Planck's constant. The experimental results for lower laser intensities are reflected back reasonably well in our theoretical calculations, both in the low-enegy and in the high-energy part of the spectra. On the basis of these investigation we conclude that both our experimental and theoretical results support the physical picture offered by our concept of plasmon-induced oscillating near field (see [14-16], [7] and [17]) of a double layer at the metal-vacuum interface.

**Acknowledgements.** This work has been supported by the Hungarian National Scientific Research Foundation OTKA, Grant No. K73728. The authors thank the Referees for their valuable comments, which helped in improving the quality of the final version of the paper.

# Figures

**Figure 1a**

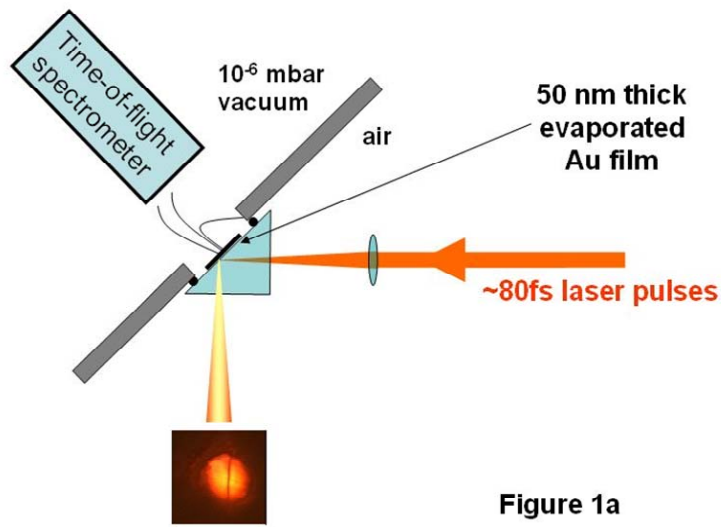

**Figure 1b**

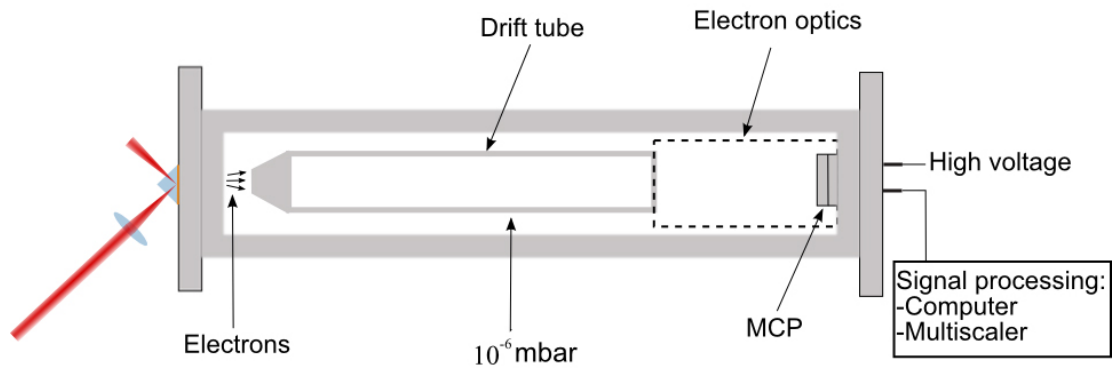

**Figure 2**

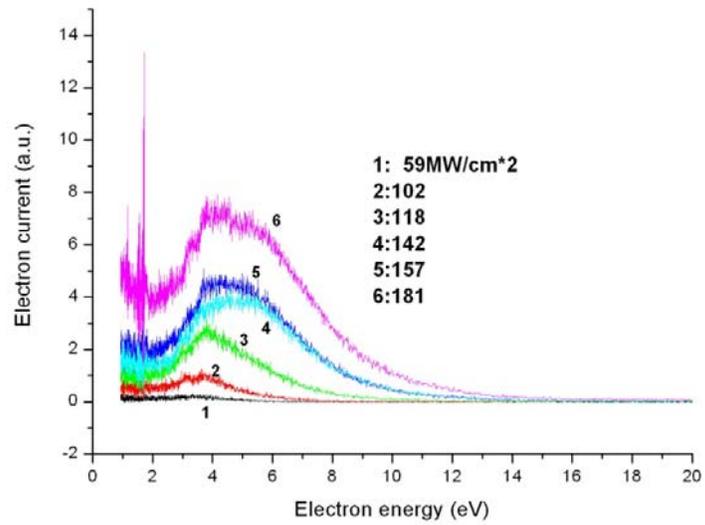

**Figure 3a-b-c-d**

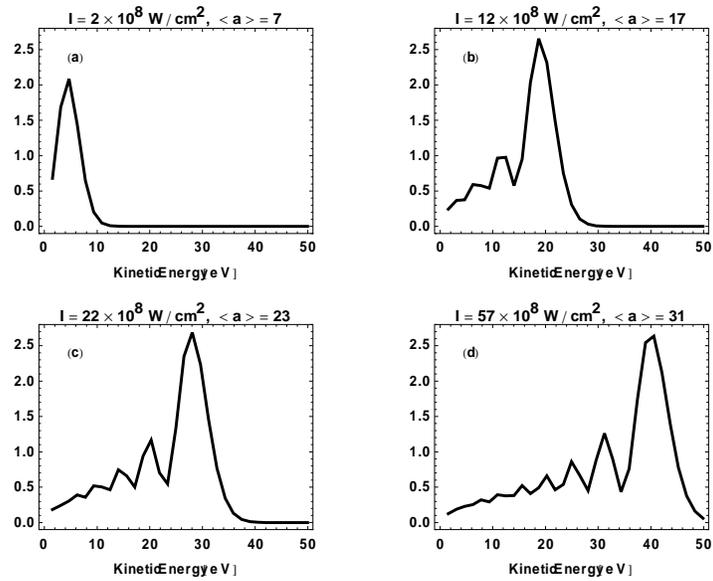

Figure 4

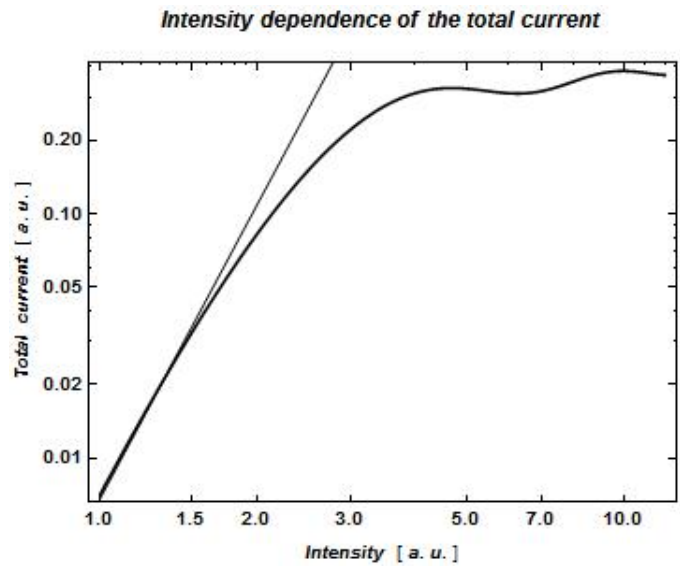

# Figure captions

**Figure 1.** Lay-out of the experiment (1.a.) and of the time-of-flight spectrometer (1.b.)

**Figure 2.** Shows typical examples of the SPO-assisted electron spectra for relatively modest laser intensities (59, 102, 118, 142, 157 and 181 MW/cm$^2$). The curve labeled by "6" essentially corresponds to the theoretical plot (Fig. 3a) of the spectrum for the lowest intensity.

**Figure 3.** Shows the calculated above-threshold spectrum of the electron current ejected due to multiplasmon absorption. On the vertical axes we have plotted the normalized current, i.e. $10 \times [(n\hbar\omega_0 - W)/E_F]^{1/2} J_n^2(<a>)$ for $n = n_0, n_0 +1, ... = 4, 5, 6, ...,$ where $n_0 = 4$ is the minimum number of photons (plasmons) for the ejection. The values of the relative current components have been evaluated by taking the average of three neighboring values associated to the multiplasmon indeces $n-1$, $n$ and $n+1$. The discrete points are connected by thick lines in order to guide the eye. The parameter $<a> = 2V_D / \hbar\omega_0$ is the ratio of the total jump of the plasmon-induced near-field potential at the metal-vacuum interface to the photon energy of the incoming laser radiation. The assumed intensities $I = \{2.7, 6.2, 12.7, 23.1\} \times 10^8 W/cm^2$ of the laser and the associated $a$-values $<a> = \{3.5, 5.3, 12.7, 20.5\}$ are separately displayed in the plot labels of figures (a), (b), (c) and (d), respectively.

**Figure 4.** Shows the intensity dependence of the total photocurrent on a log-log plot, approximated only by the Bessel factors. The thin line represent the perturbative 4th power dependence $j \propto I^4$ (the function f(x)=x$^4$/145 has been drawn for comparison). The thick line shows the non-perturbative result, which has been approximated by

$$j \sim \sum_{n=5}^{\infty} J_n^2(a) = (1/2)[1 - J_0^2(a)] - [J_1^2(a) + J_2^2(a) + J_3^2(a) + J_4^2(a)],$$ merely for illustrative

purposes. Here $a = 3.5 \times \sqrt{x}$ has been taken, where the variable $x$ is the dimensionless relative intensity (normalized to the smalles intensity value $2.7 \times 10^8 W/cm^2$).